\documentclass[sigconf,nonacm]{acmart}

\usepackage{booktabs}
\usepackage{multirow}
\usepackage{amsmath}

\usepackage{amssymb}
\usepackage{graphicx}
\graphicspath{{figures/}}
\usepackage{xcolor}
\usepackage{microtype}
\usepackage{cleveref}
\usepackage{array}
\usepackage{tabularx}
\usepackage{siunitx}

\definecolor{cooperativeblue}{RGB}{31,119,180}
\definecolor{aggressivered}{RGB}{214,39,40}
\definecolor{neutralgreen}{RGB}{44,160,44}

\newcommand{\PA}{P_{\mathrm{A}}}
\newcommand{\PC}{P_{\mathrm{C}}}
\newcommand{\PN}{P_{\mathrm{N}}}
\newcommand{\Dnoise}{\Delta_{\mathrm{noise}}}
\newcommand{\ICD}{\ensuremath{\mathrm{ICD}}}
\newcommand{\IPD}{\textsc{ipd}}
\newcommand{\LLM}{\textsc{llm}}
\newcommand{\MAS}{\textsc{mas}}

\title{Evolutionary Dynamics of Cooperation in\\
       Next-Generation LLM Agent Systems:\\
       A Cross-Provider Empirical Extension}

\author{Francisco León Zúñiga Bolívar}
\affiliation{%
  \institution{Institución Universitaria Colegio Mayor del Cauca}
  \city{Popayán}
  \country{Colombia}}
\email{franciscoleon@unimayor.edu.co}

\begin{document}

\begin{abstract}
Do next-generation \LLM agents inherit the cooperative biases documented in
their predecessors, or does scale and provider diversity reshape equilibrium
behaviour in competitive multi-agent settings? Willis et al.\ \cite{Willis2025_llm_ipd}
established a benchmark for this question using evolutionary game theory and
the Iterated Prisoner's Dilemma (\IPD), finding consistent cooperative biases
in ChatGPT-4o and Claude~3.5~Sonnet. We extend this benchmark to four frontier
models released in 2025--2026---\textbf{Claude~Sonnet~4.6}, \textbf{Gemini~2.5~Flash},
\textbf{Gemini~3.1~Pro}, and \textbf{GPT-5.4~Mini}---applying the identical
protocol across three prompting styles (Default, Prose, Self-Refine) and four
population compositions (balanced and biased, with and without noise).
Cooperative bias persists across providers (\textbf{H1}): ten of twelve
model--prompt combinations favour cooperative equilibria in balanced noiseless
conditions. Cross-provider divergence is substantial (\textbf{H3}): Gemini~2.5~Flash
reaches up to 77\% aggressive equilibria under biased conditions, while GPT-5.4~Mini
reaches 70\% cooperative equilibria under Self-Refine. Support for aggressive
capability parity is partial (\textbf{H2}): Self-Refine raises \ICD in all
models and Gemini~3.1~Pro Refine achieves the highest \ICD in the dataset
(0.925), but Default and Prose prompts show no systematic narrowing. Evidence on noise robustness is directionally positive but not
robustly confirmed (\textbf{H4}): with $n{=}500$ Moran iterations per
condition, average $\Delta_{\text{noise}}$ is $\approx 6$ percentage points for
Claude~Sonnet~4.6 versus 13~pp for Claude~3.5~Sonnet, but this cross-study
gap is not statistically significant once the predecessor's unreported
sampling error is propagated. Provider identity, rather than model generation,
is the strongest correlate of equilibrium outcomes; noise remains a universal
challenge regardless of model size or vintage.

\end{abstract}


\maketitle

\section{Introduction}

When \LLM-powered agents interact repeatedly in competitive settings, do
they cooperate or defect? The question is not merely theoretical: autonomous
\LLM agents are already deployed to negotiate contracts, allocate computational
resources, and bid in markets \cite{Wang2024_survey}. In each of these settings,
the long-run social welfare of the system hinges on whether evolutionary pressure
selects for cooperative or aggressive behaviour.

\citet{Willis2025_llm_ipd} provided the first systematic treatment
of this question, using the Iterated Prisoner's Dilemma (\IPD) as a formal
testbed. Rather than prompting \LLM{}s to output individual actions---an
approach that prior work found unreliable \cite{Fan2024_rational}---they prompt
models to generate \emph{complete strategies} in natural language, which are
subsequently implemented as Python algorithms. By simulating populations of
agents through a Moran evolutionary process, they characterised the equilibrium
tendencies of ChatGPT-4o and Claude~3.5~Sonnet. Their central finding is a
persistent cooperative bias: in balanced populations and under most prompting
conditions, cooperative strategies dominate, with aggressive equilibria arising
at substantially below prior probability.

Two gaps limit the reach of that benchmark. First, the \LLM landscape has
shifted substantially since that study. A new generation of frontier models,
released throughout 2025--2026, incorporates extended context, revised alignment
training, and improved reasoning capabilities. Whether their cooperative
tendencies persist, deepen, or erode under evolutionary pressure is unknown.
Second, the original study covered only two models from two providers, leaving
it unclear whether cooperative bias is a universal property of frontier \LLM{}s
or an artefact of specific training choices. No systematic cross-provider
comparison has been conducted.

We address both gaps by applying the identical experimental protocol to four
next-generation models spanning three providers:
\begin{itemize}
  \item \textbf{Claude~Sonnet~4.6} (Anthropic) --- successor to Claude~3.5~Sonnet;
  \item \textbf{Gemini~2.5~Flash} (Google) --- high-throughput frontier model;
  \item \textbf{Gemini~3.1~Pro} (Google) --- Google's premium-tier model;
  \item \textbf{GPT-5.4~Mini} (OpenAI) --- successor to ChatGPT-4o.
\end{itemize}

We evaluate four hypotheses, stated in advance of analysis:
\begin{description}
  \item[\textbf{H1}] \emph{Cooperative bias persists.} Next-generation models
    will maintain or increase the cooperative bias documented by Willis et al.,
    reflecting continued improvements in value alignment.
  \item[\textbf{H2}] \emph{Aggressive capability parity.} Better reasoning will
    enable more effective aggressive strategies, reducing the payoff advantage
    that cooperative agents currently hold over aggressive ones.
  \item[\textbf{H3}] \emph{Cross-provider divergence.} Differences in cooperative
    tendencies across providers will be large enough to attribute to distinct
    training and alignment choices rather than sampling noise.
  \item[\textbf{H4}] \emph{Noise robustness improves.} Newer models will be
    more robust to action noise, particularly Claude, which exhibited marked
    noise sensitivity in the original benchmark.
\end{description}

Three contributions follow from this design. First, we extend the
\citet{Willis2025_llm_ipd} benchmark---its specific Moran/attitude-agent
framework---to four new frontier models and three providers, providing the
first \emph{within-lineage longitudinal} comparison under that framework
(prior cross-provider LLM-IPD work exists but does not track generational
successors within a provider); H1 is confirmed in 10 of 12 model--prompt
combinations, establishing cooperative bias as a broadly shared property of
current frontier \LLM{}s. Second, we introduce the Index of Differential
Capabilities (\ICD), a scalar summary of the aggressive--cooperative payoff gap;
across conditions, \ICD ranges from 0.454 to 0.925, with Gemini~3.1~Pro under
Self-Refine prompting reaching the highest value in the dataset. Third,
we provide a longitudinal comparison within the Anthropic and OpenAI lineages;
provider identity is the strongest \emph{correlate} of equilibrium outcome---Gemini~2.5
Flash produces aggressive equilibria in up to 77\% of biased-population runs,
while GPT-5.4~Mini reaches 70\% cooperative equilibria under Refine prompting,
confirming H3. All Moran results use $n{=}500$ iterations per
condition; this resolves the $n{=}100$ underpowering of H4, though the
remaining cross-study comparison to the predecessor leaves H4 directionally
suggestive rather than robustly confirmed.

The remainder of this paper is organised as follows.
Section~\ref{sec:related} reviews related work.
Section~\ref{sec:method} summarises the experimental protocol.
Section~\ref{sec:results} presents our findings.
Section~\ref{sec:discussion} evaluates the four hypotheses and discusses
implications for \MAS design.
Section~\ref{sec:conclusion} concludes.

\section{Related Work}
\label{sec:related}

\paragraph{LLMs in game-theoretic settings.}
The intersection of \LLM{}s and game theory has developed rapidly into a
coherent subfield \citep{Sun2025_survey}.
\citet{Aher2023_using} used \LLM{}s to replicate human subject behaviour in
behavioural economics experiments. \citet{Brookins2023_playing} and
\citet{Guo2023_gpt} investigated whether \LLM{}s approximate Nash-rational play,
finding mixed evidence across game types. \citet{Akata2025_repeated}
studied repeated normal-form games, while \citet{Fan2024_rational} demonstrated
systematic failures when \LLM{}s are prompted to output individual game actions.
These limitations motivate our strategy-generation approach, which operates at a
higher level of abstraction and separates policy specification from action
execution.

\paragraph{LLMs in social dilemmas.}
\citet{Yocum2023_mitigating} and \citet{Piatti2024_cooperate}
studied \LLM agent behaviour in Markov social dilemmas. \citet{Park2023_generative}
introduced generative agents for simulating social behaviour.
\citet{DeZarza2023_emergent} explored emergent cooperation in
\LLM-extended coevolutionary theory. \citet{Leibo2017_multiagent} grounded such
studies in multi-agent reinforcement learning. Our contribution to this strand is
the analysis of \emph{evolutionary population dynamics}---how strategy types rise
or collapse across generations---rather than individual agent decisions at a
single time step.

\paragraph{The Willis et al.\ (2025) benchmark.}
\citet{Willis2025_llm_ipd} introduced the framework we extend. They prompt
\LLM{}s to generate natural-language \IPD strategies (converted to Python), then
simulate Moran evolutionary processes over populations of attitude-agents
(Aggressive, Cooperative, Neutral). Key findings include a cooperative bias in
both ChatGPT-4o and Claude~3.5~Sonnet, with Claude exhibiting greater difficulty
generating effective aggressive strategies and notable fragility under action
noise. Applying the Self-Refine framework \citep{Madaan2023_self_refine} to strategy
generation narrowed this gap \citep{Willis2025_llm_ipd}, demonstrating that
iterative self-critique can inadvertently enable aggressive strategies, a result
that complicates alignment-by-design assumptions.

\paragraph{Concurrent and subsequent work.}
Three papers have since tested related questions with different frameworks.
\citet{Payne2025_strategic} run evolutionary \IPD tournaments across GPT, Gemini,
and Claude frontier models using variable termination probabilities and 32,000
prose rationales, without a Moran process or attitude-agent populations; they
find the same provider fingerprint we document (Gemini aggressive, OpenAI
cooperative), providing independent corroboration for our cross-provider
divergence finding (H3, Section~\ref{sec:discussion}) via a mechanistically
distinct design. \citet{Vallinder2024_cultural} apply cross-provider evolutionary
selection to the Donor Game under indirect reciprocity with older models
(Claude~3.5~Sonnet, Gemini~1.5~Flash, GPT-4o) and find the ranking
Claude $>$ Gemini $>$ GPT; our GPT-5.4~Mini emerges as the most cooperative of
the four models we test, inverting that OpenAI position and suggesting that model
generation, not just provider, shapes cooperative disposition.
\citet{Willis2026_collective} scale the Willis et al.\ framework to hundreds of
agents under cultural evolutionary dynamics and find that newer \LLM{}s
\emph{worsen} societal outcomes in their setting---an apparent tension with our
result that cooperative bias persists under Moran selection at $n{=}12$. Whether
the selection mechanism (pairwise Moran versus cultural evolution) or population
scale reconciles the divergence is an open question we flag for future work.

\paragraph{Cross-generational and cross-provider comparisons.}
Prior capability comparisons across \LLM providers focused on general reasoning
benchmarks---MMLU \citep{Hendrycks2021_mmlu}, HumanEval \citep{Chen2021_humaneval},
and similar---where cooperative dispositions are absent by design. The work above establishes that provider-level differences emerge in
evolutionary social dilemmas, but each study either uses older model generations
or a single dilemma type. Within-lineage longitudinal analysis---whether
successive generations from the same provider preserve or reverse established
cooperation patterns---remains absent from the literature. Our study targets that
gap by pairing current frontier models against their immediate predecessors under
the same Moran \IPD framework.

\paragraph{Evolutionary game theory foundations.}
Our simulations rest on the Moran process \citep{Moran1958_random}, the classical
model of evolutionary selection in finite populations; \citet{Nowak2004_cooperation}
established its connection to cooperation in \IPD games in finite populations, and
\citet{Traulsen2006_fixation} derived fixation probabilities under this process.
The \IPD framework follows \citet{Axelrod1984_evolution,Axelrod1981_evolution};
\citet{Nowak2006_evolutionary} provides the backbone for convergence analysis in
finite evolutionary games. The noise mechanism follows
\citet{WuAxelrod1995_noise} and \citet{WahlNowak1999_noise}.
These threads---\LLM strategic behaviour, multi-agent social dilemmas, and
evolutionary population dynamics---jointly define the space this paper occupies:
a cross-provider benchmark for next-generation \LLM{}s, including within-lineage
longitudinal comparisons, evaluated under Darwinian selection over \IPD strategy
populations.

\section{Method}
\label{sec:method}

We apply the identical experimental protocol of Willis et al.\ \cite{Willis2025_llm_ipd}
to four new models, adding the first Google provider entry and enabling
cross-provider comparison. We summarise the key components below.

\subsection{Strategy Generation}

For each combination of model and prompt style, we generate 25 natural-language
\IPD strategies per attitude (Aggressive, Cooperative, Neutral), yielding 75
strategies per model--prompt pair. Three prompt styles are used (Table~\ref{tab:prompts}):
\textbf{Default} (direct elicitation with game-theoretic language),
\textbf{Refine} (Self-Refine \cite{Madaan2023_self_refine} applied to the default
output), and \textbf{Prose} (obfuscated scenario without game-theoretic terminology).

\begin{table}[h]
\caption{Prompt styles (following Willis et al.\ \cite{Willis2025_llm_ipd}).}
\label{tab:prompts}
\small
\begin{tabular}{@{}lp{5.5cm}@{}}
\toprule
\textbf{Style} & \textbf{Description} \\
\midrule
Default & Direct prompt with game-theoretic language; strategy generated in natural language. \\
Refine  & Default output refined via Self-Refine \cite{Madaan2023_self_refine}: the model critiques and rewrites its own strategy. \\
Prose   & Game-theoretic framing obfuscated as a real-world scenario (e.g., trade negotiation). Strategy generated, then translated to the \IPD context. \\
\bottomrule
\end{tabular}
\end{table}

Strategies are then converted to Python algorithms using the same \LLM that
generated them (one per provider), preserving consistency within each provider.
Strategies that cannot be executed without error are replaced by regeneration
until 25 valid strategies per attitude are obtained.

\subsection{Models}

We apply the protocol to the four models listed in Table~\ref{tab:models}.
These represent the direct successors of the models studied by Willis et al.\
(Claude lineage, OpenAI lineage) and a new provider (Google).

\begin{table}[h]
\caption{Models evaluated in this study.}
\label{tab:models}
\small
\begin{tabular}{@{}llll@{}}
\toprule
\textbf{Model} & \textbf{Provider} & \textbf{Predecessor} & \textbf{Released} \\
\midrule
Claude Sonnet 4.6    & Anthropic & Claude 3.5 Sonnet  & 2025 \\
Gemini 2.5 Flash     & Google    & (new provider)     & 2025 \\
Gemini 3.1 Pro       & Google    & (new provider)     & 2025 \\
GPT-5.4 Mini         & OpenAI    & ChatGPT-4o         & 2025 \\
\bottomrule
\end{tabular}
\end{table}

\subsection{IPD Tournament}

All 75 strategies compete in an all-play-all tournament using the Axelrod Python
library \cite{Knight2016_open}. Each match consists of 1,000 rounds of the
standard \IPD (payoff matrix: $R=3$, $S=0$, $T=5$, $P=1$). Noise conditions
introduce a 10\% probability of action-flip per player per round. Tournaments
are repeated 20 times.

\subsection{Attitude-Agents}

Following Willis et al.\ \cite{Willis2025_llm_ipd}, we define three attitude-agents,
each uniformly sampling from its corresponding strategy set for each match.
This captures populations of agents with distinct strategic dispositions rather
than fixed individual strategies.

\subsection{Moran Process}

We simulate Moran evolutionary processes with population size $n=12$ and 500
iterations per condition. Four population compositions are evaluated:
\begin{enumerate}
  \item \textbf{Balanced, clean} (4:4:4) --- equal priors;
  \item \textbf{Biased, clean} (8:2:2) --- aggressive majority;
  \item \textbf{Balanced, noise} (4:4:4 with noise) --- equal priors with action noise;
  \item \textbf{Biased, noise} (8:2:2 with noise) --- aggressive majority with action noise.
\end{enumerate}

\begin{figure}[h]
  \centering
  \includegraphics[width=\columnwidth]{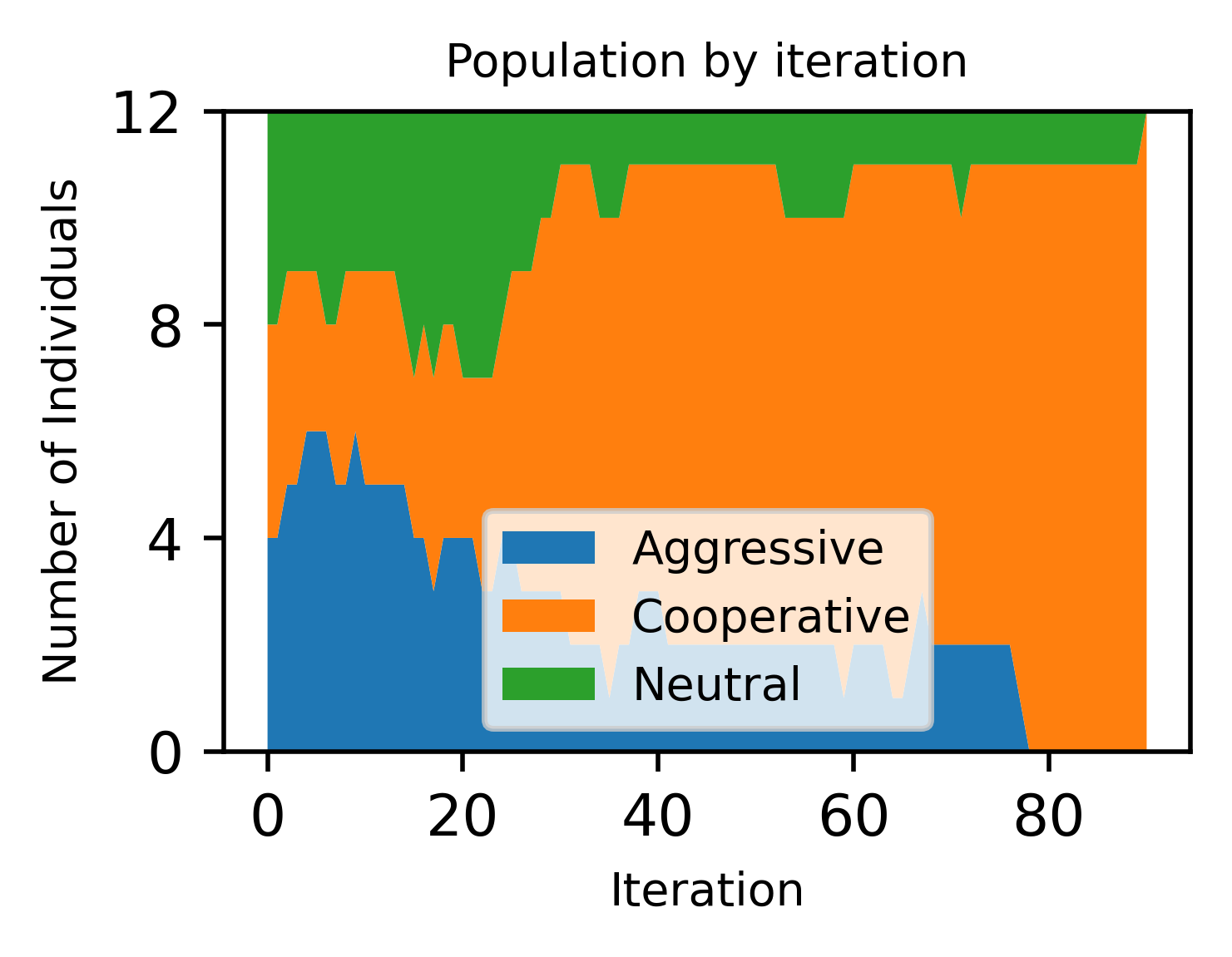}
  \Description{Line chart showing population fractions over evolutionary
               generations for three attitude groups (Aggressive, Cooperative,
               Neutral). Cooperative fraction rises to fixation.}
  \caption{Example Moran process trajectory for Claude~4.6~Default
           (balanced 4:4:4 population, single seed).
           Shading represents the fraction of agents with each attitude
           at each generation; the process terminates when one attitude
           reaches fixation.}
  \label{fig:moran_traj}
\end{figure}

Convergence is assessed by the proportion of runs reaching each monoculture
equilibrium (all-Aggressive, all-Cooperative, or all-Neutral).
Figure~\ref{fig:moran_traj} illustrates a single trajectory.

\subsection{Derived Metrics}

We introduce the \emph{Index of Differential Capabilities} (\ICD) to summarise
the head-to-head payoff gap between aggressive and cooperative agents:
\begin{equation}
  \ICD = \frac{\bar{u}(A)}{\bar{u}(C)}, \quad
  \bar{u}(k) = \frac{1}{3}\sum_{j \in \{A,C,N\}} u(k,j),
  \label{eq:icd}
\end{equation}
where $u(k,j)$ is the normalised payoff of attitude $k$ against attitude $j$.
An \ICD of 1.0 implies equal capability; values below 1.0 indicate a cooperative
advantage. We also define the \emph{noise sensitivity} $\Dnoise$:
\begin{equation}
  \Dnoise = \PC^{\text{clean}} - \PC^{\text{noise}},
  \label{eq:dnoise}
\end{equation}
measuring the drop in cooperative equilibrium probability under action noise.

\section{Results}
\label{sec:results}

\subsection{Strategy Validation: Cooperation Propensity}

Table~\ref{tab:cooperation} reports the normalised propensity to cooperate
for the Default prompt without noise, analogous to Table~3 in Willis et al.

\begin{table}[h]
\caption{Normalised cooperation propensity (Default prompt, no noise).
         Rows: row-player attitude; columns: opponent attitude.}
\label{tab:cooperation}
\small
\setlength{\tabcolsep}{4pt}
\begin{tabular}{@{}llccc@{}}
\toprule
\textbf{Model} & \textbf{Att.} & \textbf{vs A} & \textbf{vs C} & \textbf{vs N} \\
\midrule
\multirow{3}{*}{Claude 4.6}
  & A & 0.000 & 0.297 & 0.286 \\
  & C & 0.298 & 0.991 & 0.992 \\
  & N & 0.287 & 0.993 & 1.000 \\
\midrule
\multirow{3}{*}{Gemini 2.5 Flash}
  & A & 0.082 & 0.087 & 0.087 \\
  & C & 0.088 & 1.000 & 1.000 \\
  & N & 0.088 & 1.000 & 1.000 \\
\midrule
\multirow{3}{*}{Gemini 3.1 Pro}
  & A & 0.413 & 0.629 & 0.629 \\
  & C & 0.630 & 0.999 & 0.999 \\
  & N & 0.630 & 0.999 & 0.999 \\
\midrule
\multirow{3}{*}{GPT-5.4 Mini}
  & A & 0.006 & 0.025 & 0.023 \\
  & C & 0.026 & 1.000 & 0.940 \\
  & N & 0.024 & 0.940 & 0.883 \\
\bottomrule
\end{tabular}
\end{table}

All four models exhibit the expected attitude separation: aggressive strategies
cooperate markedly less than cooperative or neutral strategies.
Gemini~3.1~Pro is the notable exception, with aggressive strategies cooperating
at 41\% even against other aggressive strategies---substantially higher than any
model tested by Willis et al. By contrast, Gemini~2.5~Flash and GPT-5.4~Mini
generate highly committed aggressive strategies (cooperation rates near zero
against all opponents), while Claude~4.6 closely mirrors Claude~3.5~Sonnet's
pattern: aggressive strategies cooperate at $\approx 0$\% against other
aggressors but rise to $\approx 29$\% against cooperative opponents.

\subsection{Head-to-Head Payoffs and Differential Capabilities}

Table~\ref{tab:payoffs} presents the normalised mean payoffs across all
25 matches for each attitude pairing (Default prompt, no noise), together with
the \ICD (Eq.~\ref{eq:icd}).

\begin{table*}[t]
\caption{Normalised head-to-head payoffs (Default prompt, no noise) and
         Index of Differential Capabilities (\ICD). Lower \ICD indicates
         a larger cooperative advantage.}
\label{tab:payoffs}
\small
\begin{tabular}{@{}llccccccc@{}}
\toprule
\textbf{Model} & \textbf{Prompt} &
  \multicolumn{3}{c}{\textbf{Aggressive payoff vs.}} &
  \multicolumn{3}{c}{\textbf{Cooperative payoff vs.}} &
  \textbf{\ICD} \\
\cmidrule(lr){3-5}\cmidrule(lr){6-8}
 &  & A & C & N & A & C & N & \\
\midrule
\multirow{3}{*}{Claude 4.6}
  & Default & 1.000 & 1.893 & 1.861 & 1.890 & 2.982 & 2.988 & 0.605 \\
  & Prose   & 1.242 & 1.917 & 1.784 & 1.507 & 3.000 & 2.836 & 0.673 \\
  & Refine  & 2.432 & 2.820 & 2.683 & 2.772 & 2.997 & 2.924 & 0.913 \\
\midrule
\multirow{3}{*}{Gemini 2.5 Flash}
  & Default & 1.232 & 1.244 & 1.244 & 1.239 & 3.000 & 3.000 & 0.514 \\
  & Prose   & 1.290 & 1.796 & 1.551 & 0.985 & 2.574 & 2.086 & 0.822 \\
  & Refine  & 1.486 & 2.288 & 2.022 & 1.470 & 2.998 & 2.635 & 0.816 \\
\midrule
\multirow{3}{*}{Gemini 3.1 Pro}
  & Default & 1.992 & 2.400 & 2.400 & 2.397 & 2.998 & 2.998 & 0.809 \\
  & Prose   & 2.009 & 1.971 & 1.972 & 1.702 & 2.979 & 2.858 & 0.789 \\
  & Refine  & 2.159 & 2.405 & 2.433 & 2.259 & 2.685 & 2.617 & 0.925 \\
\midrule
\multirow{3}{*}{GPT-5.4 Mini}
  & Default & 1.017 & 1.079 & 1.073 & 1.074 & 3.000 & 2.900 & 0.454 \\
  & Prose   & 1.982 & 2.395 & 2.423 & 2.394 & 2.914 & 2.937 & 0.825 \\
  & Refine  & 1.066 & 1.427 & 1.252 & 1.343 & 2.846 & 2.305 & 0.577 \\
\bottomrule
\end{tabular}
\end{table*}

\begin{figure}[h]
  \centering
  \includegraphics[width=\columnwidth]{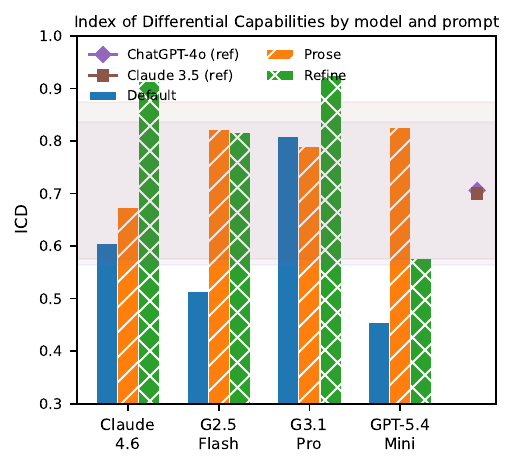}
  \Description{Grouped bar chart showing ICD values for four LLM models
               (Claude 4.6, Gemini 2.5 Flash, Gemini 3.1 Pro, GPT-5.4 Mini)
               across three prompt styles (Default, Prose, Refine).
               Values range from 0.454 to 0.925.}
  \caption{Index of Differential Capabilities (\ICD) per model and prompt style.
           Lower values indicate a larger cooperative payoff advantage.
           Shaded bands show the range of \ICD values reported by Willis et al.\
           for ChatGPT-4o (purple) and Claude~3.5~Sonnet (brown); markers indicate
           per-model averages.}
  \label{fig:icd}
\end{figure}

\ICD values range from 0.454 (GPT-5.4~Mini Default) to 0.925 (Gemini~3.1~Pro~Refine)
and vary substantially across prompts. Cooperative and neutral attitudes achieve
near-mutual-cooperation payoffs ($\approx 3.0$) in most model--prompt combinations,
consistent with the original findings. GPT-5.4~Mini Default's \ICD of 0.454
means aggressive strategies achieve less than half the payoff of cooperative ones;
at the other extreme, Gemini~3.1~Pro~Refine (0.925) and Claude~4.6~Refine (0.913)
approach payoff parity. The Self-Refine prompt increases \ICD relative to Default in all four models,
replicating the original finding that Self-Refine narrows the aggressive--cooperative
gap. For Claude~4.6~Refine, this effect is particularly pronounced: aggressive
strategies achieve payoffs of 2.43--2.82, approaching cooperative strategies at
2.77--3.00.

One notable exception to the monotonic Default$\to$Prose$\to$Refine trend is
GPT-5.4~Mini: its Refine \ICD (0.577) is substantially \emph{lower} than its
Prose \ICD (0.825), the only reversal in the dataset. This suggests that
GPT-5.4~Mini's Self-Refine mechanism improves strategy coherence at the expense
of aggressive payoff, producing more cooperative-dominated refinements rather
than capability-balanced ones.

Compared to Willis et al.'s Claude~3.5~Sonnet (Table~4 therein), Claude~4.6
shows systematically higher payoffs for aggressive strategies under Refine,
suggesting improved aggressive strategy generation in the newer version.
GPT-5.4~Mini Default, however, shows weaker aggressive performance than
ChatGPT-4o~Default (1.05 vs.\ 1.81 average), indicating that model scaling
does not uniformly improve aggressive capability.

\subsection{Evolutionary Equilibria}

Table~\ref{tab:moran} reports the main Moran process results: the proportion
of 500 runs converging to each equilibrium attitude, for all four population
conditions. The Willis et al.\ reference values are shown for comparison.

\begin{table*}[t]
\caption{Moran process equilibrium proportions (\%A / \%C / \%N) across
         four population conditions. Prior probability indicates the
         baseline for each initial ratio.
         Reference values from Willis et al.\ \cite{Willis2025_llm_ipd}
         (Table~6) are shown in the bottom rows.
         Bold: values highlighted in the text discussion.}
\label{tab:moran}
\footnotesize
\setlength{\tabcolsep}{3pt}
\begin{tabular}{@{}llcccc@{}}
\toprule
\textbf{Model} & \textbf{Prompt} &
  \textbf{4:4:4 clean} &
  \textbf{4:4:4 noise} &
  \textbf{8:2:2 clean} &
  \textbf{8:2:2 noise} \\
\cmidrule(lr){3-3}\cmidrule(lr){4-4}\cmidrule(lr){5-5}\cmidrule(lr){6-6}
 & & (prior: 33/33/33) & (prior: 33/33/33) & (prior: 67/17/17) & (prior: 67/17/17) \\
\midrule
\multirow{3}{*}{Claude 4.6}
  & Default & 2/49/\textbf{49}  & 24/\textbf{40}/36 & 12/\textbf{47}/41 & 60/23/17 \\
  & Prose   & 5/\textbf{48}/47  & 28/37/36          & 32/38/30          & 64/17/19 \\
  & Refine  & 23/\textbf{39}/38 & 29/\textbf{41}/30 & 59/22/20          & 62/23/15 \\
\midrule
\multirow{3}{*}{Gemini 2.5 Flash}
  & Default & 2/\textbf{49}/48  & 20/39/\textbf{41} & 24/38/38          & 61/23/16 \\
  & Prose   & 24/\textbf{47}/30 & 40/28/32          & \textbf{65}/19/16 & \textbf{77}/9/14 \\
  & Refine  & 21/\textbf{47}/32 & 36/32/32          & 56/22/22          & 73/13/13 \\
\midrule
\multirow{3}{*}{Gemini 3.1 Pro}
  & Default & 14/42/\textbf{45} & 31/35/34 & 36/32/32          & 63/20/17 \\
  & Prose   & 19/\textbf{43}/38 & 36/29/34 & 60/21/20          & 70/12/17 \\
  & Refine  & 27/\textbf{40}/33 & 32/37/31 & 62/18/20          & 66/18/16 \\
\midrule
\multirow{3}{*}{GPT-5.4 Mini}
  & Default & 2/\textbf{53}/45          & 14/\textbf{47}/39 & 22/38/\textbf{40} & 52/25/23 \\
  & Prose   & 12/\textbf{45}/43         & 31/34/35          & 46/28/26          & 66/16/18 \\
  & Refine  & 4/\textbf{70}/26          & 25/\textbf{42}/33 & 28/\textbf{46}/26 & 60/19/20 \\
\midrule
\midrule
\multirow{3}{*}{\textit{ChatGPT-4o}$^\dagger$}
  & Default & 14/53/33 & 16/42/42 & 66/19/17 & 59/20/21 \\
  & Prose   & 13/38/49 & 23/41/36 & 35/27/38 & 60/18/22 \\
  & Refine  & 19/48/33 & 28/38/34 & 49/30/21 & 63/19/18 \\
\multirow{3}{*}{\textit{Claude 3.5 Sonnet}$^\dagger$}
  & Default & 4/49/47  & 15/37/48 & 36/24/40 & 41/20/39 \\
  & Prose   & 14/42/44 & 17/33/50 & 41/30/29 & 61/26/13 \\
  & Refine  & 16/51/33 & 37/34/29 & 50/22/28 & 60/18/22 \\
\bottomrule
\end{tabular}
\vspace{2pt}\\
\small$^\dagger$From Willis et al.\ \cite{Willis2025_llm_ipd}, Table~6.
\end{table*}

\begin{figure*}[t]
  \centering
  \includegraphics[width=\textwidth]{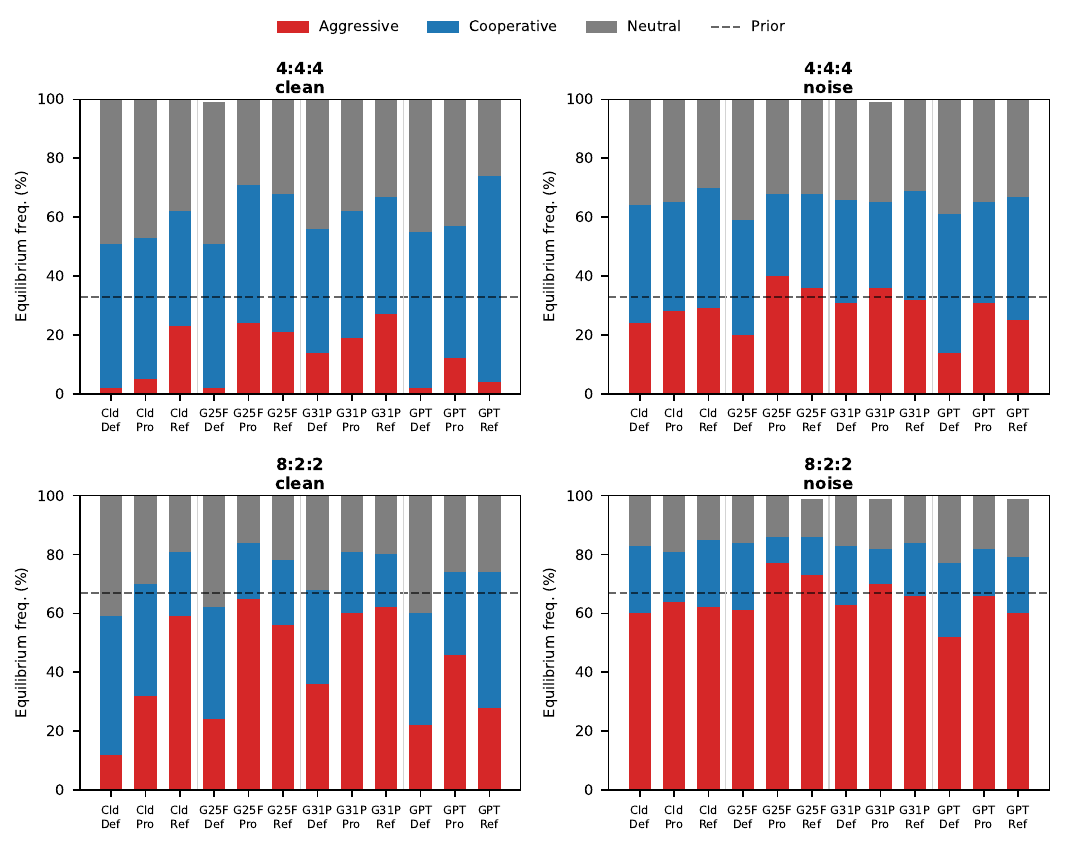}
  \Description{Four-panel stacked bar chart (balanced clean, balanced noisy,
               biased clean, biased noisy) showing Aggressive, Cooperative,
               and Neutral equilibrium proportions for all 12 model-prompt
               combinations. Cooperative dominates balanced conditions;
               Aggressive rises under noise and bias.}
  \caption{Moran process equilibrium distributions across all 48 conditions
           (12 model--prompt combinations $\times$ 4 population regimes).
           Stacked bars show the proportion of 500 runs converging to
           Aggressive (red), Cooperative (blue), or Neutral (grey) equilibrium.
           Dashed line indicates the theoretical prior (33\% for balanced,
           67\% for biased conditions). Vertical grey lines separate the
           four LLM families.}
  \label{fig:equilibria}
\end{figure*}

\paragraph{Balanced, noiseless condition (4:4:4 clean).}
This condition provides the most direct comparison with the original benchmark.
Ten of twelve model--prompt combinations favour cooperative equilibria as
the plurality outcome ($\PC > \PA$ and $\PC > \PN$), confirming a widespread
cooperative bias. The strongest cooperative equilibria are achieved
by GPT-5.4~Mini~Refine (70\%), GPT-5.4~Mini~Default (53\%), and
Gemini~2.5~Flash~Default (49\%).
The two exceptions are both cases where the \emph{Neutral} attitude achieves the
plurality: Claude~4.6~Default (2\%A/49\%C/\textbf{49\%N}, a near-tie resolved
in favour of Neutral by 4 of 500 runs) and Gemini~3.1~Pro~Default
(14\%A/42\%C/\textbf{45\%N}). Notably, neither exception is driven by
aggressive dominance; in both cases aggressive equilibria remain well below
the 33\% prior. The Neutral plurality for Gemini~3.1~Pro~Default likely
reflects that its aggressive strategies cooperate at unusually high rates
(Table~\ref{tab:cooperation}, 41\% vs.\ own attitude), blurring the boundary
between aggressive and neutral behaviour and allowing neutral strategies to
benefit from mutual cooperation without defection penalties.
The larger $n{=}500$ sample, together with the corrected complete strategy
libraries (see the erratum note), resolves several earlier ambiguities:
GPT-5.4~Mini~Default, Gemini~3.1~Pro~Prose, and Gemini~3.1~Pro~Refine all
favour a clear cooperative plurality (53\%C, 43\%C, and 40\%C respectively).

Comparing within-lineage successors:
Claude~4.6~Default achieves 2\%A/49\%C/49\%N, virtually identical to
Claude~3.5~Sonnet~Default at 4\%A/49\%C/47\%N. This is the closest
generational replication in the dataset.
GPT-5.4~Mini~Default achieves 2\%A/53\%C/45\%N, far lower aggressive than
ChatGPT-4o~Default (14\%A), confirming improved cooperative bias in the
OpenAI lineage for this condition.

\paragraph{Balanced, noisy condition (4:4:4 noise).}
Noise consistently shifts equilibria toward aggression across all twelve
model--prompt combinations; all twelve pairs show
$\PA^{\text{noise}} > \PA^{\text{clean}}$. With $n{=}500$, the apparent
$n{=}100$ anomaly in which Gemini~3.1~Pro~Default's cooperative equilibria
rose under noise disappears: its $\Dnoise$ is now $+6$pp (normal degradation),
confirming that the earlier negative value was sampling noise.
The most noise-resilient combination is Claude~4.6~Refine, whose cooperative
equilibrium is essentially flat under noise ($\Dnoise = -1$pp; see
Section~\ref{sec:noise} and Table~\ref{tab:noise}).
Gemini~2.5~Flash~Prose shows the largest cooperative degradation
(47\% to 28\%, $\Dnoise = 19$pp), and GPT-5.4~Mini~Refine the largest overall
($\Dnoise = 28$pp).

\paragraph{Biased, noiseless condition (8:2:2 clean).}
With aggressive agents as the initial majority, the prior for aggressive
equilibria is 67\%. Below this threshold indicates resistance to aggressive
takeover; above it indicates susceptibility. At $n{=}500$, \emph{all twelve}
model--prompt combinations finish below the 67\% prior, indicating that
cooperative/neutral attitudes resist aggressive takeover even when seeded as
the minority. The most resistant are Claude~4.6~Default (12\%A) and
GPT-5.4~Mini~Default (22\%A). The closest to the prior is
Gemini~2.5~Flash~Prose at 65\%A---just below the seeding proportion,
suggesting aggressive agents enjoy roughly no disadvantage in that condition
but do not amplify beyond it.

\paragraph{Biased, noisy condition (8:2:2 noise).}
In the most adverse condition (aggressive majority with noise), all twelve
model--prompt combinations converge to aggressive equilibria at rates of
52--77\%. Contrary to a naive reading of the aggressive-majority prior (67\%),
only 2 of 12 combinations exceed this threshold---Gemini~2.5~Flash~Prose
(77\%) and Gemini~2.5~Flash~Refine (73\%). The remaining 10 of 12 fall at or
below the prior, indicating that the Moran process does not amplify aggressive
dominance beyond the initial seeding for most model--prompt combinations, even
in this worst-case scenario. The most resistant are GPT-5.4~Mini~Default (52\%)
and Claude~4.6~Default (60\%), which finish substantially below the prior.

\subsection{Cross-Provider Significance Tests}
\label{sec:ztests}

Table~\ref{tab:ztests} reports the pairwise two-sample $z$-tests for
$\PA$ in the balanced noiseless condition (Default prompt), with
Holm-Bonferroni correction for the six simultaneous comparisons.

\begin{table}[h]
\caption{Pairwise $z$-tests for aggressive equilibrium proportion $\PA$
         (balanced 4:4:4, noiseless, Default prompt; $n=500$ per condition).
         Holm-Bonferroni adjusted thresholds $\alpha_\text{adj}$ shown;
         $^{***}$: $p < \alpha_\text{adj}$, ns: not significant.}
\label{tab:ztests}
\small
\begin{tabular}{@{}llcc@{}}
\toprule
\textbf{Model A} & \textbf{Model B} & $z$ & \textbf{Sig.} \\
\midrule
G3.1 Pro (14\%)   & GPT (2\%)        & $+7.24$ & $^{***}$ \\
Claude (2\%)      & G3.1 Pro (14\%)  & $-6.92$ & $^{***}$ \\
G2.5 Flash (2\%)  & G3.1 Pro (14\%)  & $-6.61$ & $^{***}$ \\
G2.5 Flash (2\%)  & GPT (2\%)        &  $+0.90$ & ns \\
Claude (2\%)      & GPT (2\%)        &  $+0.48$ & ns \\
Claude (2\%)      & G2.5 Flash (2\%) &  $-0.43$ & ns \\
\bottomrule
\end{tabular}
\end{table}

\subsection{Strategy Diversity}
\label{sec:diversity}

To characterise how behaviourally varied the 25 strategies per attitude are,
we compute the Shannon entropy $H$ of the per-strategy cooperation rate
distribution within each attitude group, per prompt (no noise):
\begin{equation}
  H = -\sum_{i} p_i \log p_i,
\end{equation}
where $p_i$ is the fraction of strategies falling in cooperation-rate bin $i$
(10 equal bins on $[0,1]$). We also report the \emph{attitude separation},
defined as the difference in mean cooperation rate between the Cooperative
and Aggressive attitude agents.

Table~\ref{tab:diversity} summarises the results.

\begin{table}[h]
\caption{Strategy diversity: mean Shannon entropy $H$ (nats) across attitudes,
         and attitude separation (Cooperative minus Aggressive mean cooperation
         rate). Higher entropy indicates more varied within-attitude behaviour;
         higher separation indicates clearer attitude distinction.}
\label{tab:diversity}
\small
\begin{tabular}{@{}llcc@{}}
\toprule
\textbf{Model} & \textbf{Prompt} & $\bar{H}$ & \textbf{Sep.} \\
\midrule
\multirow{3}{*}{Claude 4.6}
  & Default & 0.28 & 0.56 \\
  & Prose   & 1.35 & 0.56 \\
  & Refine  & 0.68 & 0.16 \\
\midrule
\multirow{3}{*}{Gemini 2.5 Flash}
  & Default & 0.20 & 0.61 \\
  & Prose   & 1.07 & 0.41 \\
  & Refine  & 1.48 & 0.52 \\
\midrule
\multirow{3}{*}{Gemini 3.1 Pro}
  & Default & 0.65 & 0.32 \\
  & Prose   & 0.87 & 0.42 \\
  & Refine  & 1.69 & 0.17 \\
\midrule
\multirow{3}{*}{GPT-5.4 Mini}
  & Default & 0.22 & 0.63 \\
  & Prose   & 0.57 & 0.28 \\
  & Refine  & 0.62 & 0.47 \\
\bottomrule
\end{tabular}
\end{table}

Three patterns emerge. First, attitude separation for Gemini~3.1~Pro is
consistently the lowest of the four models (avg.\ 0.30 vs.\ 0.47 for the others),
consistent with its unusually high aggressive cooperation rate (Table~\ref{tab:cooperation})
and the consequent neutral-plurality outcomes in the Moran process.
When aggressive strategies cooperate almost as often as neutral ones,
the fitness differential that normally drives cooperative equilibria is eroded.
Second, the Refine prompt reduces separation in Claude~4.6 (0.16) and
Gemini~3.1~Pro (0.17), corroborating the \ICD analysis: self-refinement
narrows the aggressive--cooperative gap at the individual strategy level.
Third, Gemini~3.1~Pro~Refine attains the highest within-attitude entropy in the
dataset ($\bar{H} = 1.69$): its strategies are behaviourally varied yet weakly
separated by attitude, so high diversity \emph{within} attitudes coexists with
the low distinction \emph{between} them noted above---a combination that further
blurs the fitness signal driving the Moran dynamics.

\subsection{Noise Sensitivity ($\Dnoise$)}
\label{sec:noise}

Table~\ref{tab:noise} summarises $\Dnoise$ (Eq.~\ref{eq:dnoise})
for the balanced condition, providing a direct measure of noise robustness.

\begin{table}[h]
\caption{Noise sensitivity $\Dnoise$ (balanced population, 4:4:4).
         Positive values indicate degradation of cooperative equilibria under noise.}
\label{tab:noise}
\small
\begin{tabular}{@{}lllll@{}}
\toprule
\textbf{Model} & \textbf{Default} & \textbf{Prose} & \textbf{Refine} & \textbf{Avg.} \\
\midrule
Claude 4.6        & 9  & 12 & $-1$ & 6  \\
Gemini 2.5 Flash  & 10 & 19 & 14 & 15 \\
Gemini 3.1 Pro    & 6  & 14 & 3  & 8  \\
GPT-5.4 Mini      & 7  & 11 & 28 & 15 \\
\midrule
\textit{Claude 3.5 Sonnet}$^\dagger$ & 12 & 9 & 17 & 13 \\
\textit{ChatGPT-4o}$^\dagger$         & 11 & $-3$ & 10 & 6 \\
\bottomrule
\end{tabular}
\vspace{2pt}\\
\small$^\dagger$Derived from Willis et al.\ \cite{Willis2025_llm_ipd}, Table~6.
\end{table}

With $n{=}500$, Claude~4.6 stands out as the most noise-robust on average
($\Dnoise \approx 6$pp), with its Refine condition essentially flat under
noise ($\Dnoise = -1$pp). GPT-5.4~Mini and Gemini~2.5~Flash are the most
noise-sensitive ($\approx 15$pp each); GPT-5.4~Mini's figure is driven almost
entirely by its Refine prompt ($\Dnoise = 28$pp).
Claude~4.6's average ($\approx 6$pp) is numerically \emph{lower} than
Claude~3.5~Sonnet's 13pp. While our own estimate is now precise
(SE $\approx 2.2$pp at $n{=}500$), this is a \emph{cross-study} comparison:
the predecessor's value carries its own (unreported, likely $n{=}100$)
sampling error, and propagating it leaves the $\approx 7$pp gap short of
conventional significance. We therefore treat this as directionally suggestive
evidence of improved noise robustness rather than a confirmed effect; the full
inferential argument is given under H4 (Section~\ref{sec:discussion}).

\section{Discussion}
\label{sec:discussion}

\subsection{Hypothesis Evaluation}

\paragraph{H1 — Cooperative bias persists (\textcolor{green!60!black}{Confirmed within the tested strategy library}).}
The cooperative bias documented by Willis et al.\ is consistent across model generations
and providers. All verdicts in this section are conditional on the single
75-strategy-per-model library we generated; we did not resample alternative
strategy libraries, so the verdicts speak to these libraries' evolutionary
dynamics rather than to a population-level claim over all strategies a model
could produce. In the balanced noiseless condition---the most direct comparison
to the original benchmark---10 of 12 model--prompt combinations favour
cooperative equilibria as the plurality outcome ($\PC > \PA$ and $\PC > \PN$),
and aggressive equilibrium rates remain well below the 33\% prior for the
majority of configurations. The successor models (Claude~4.6, GPT-5.4~Mini)
show cooperative bias comparable to or stronger than their predecessors.
The within-lineage consistency is striking: Claude~4.6~Default (2\%A/49\%C)
and Claude~3.5~Sonnet~Default (4\%A/49\%C) are indistinguishable, suggesting
that the fundamental strategic disposition of Anthropic models is stable across
generations. The $n{=}500$ sample, together with the corrected complete
libraries, sharpens this picture: GPT-5.4~Mini~Default, Gemini~3.1~Pro~Prose
and Gemini~3.1~Pro~Refine all favour an unambiguous Cooperative-plurality,
bringing the count of cooperative-plurality combinations to 10 of 12.

We hypothesise that this stability reflects convergence in alignment training
objectives: both models are trained with RLHF and Constitutional AI techniques
that strongly penalise defection-like behaviours. The persistence of cooperative
bias is consistent with this interpretation but does not rule out alternative
explanations (e.g., convergence in the training data distribution for
game-theoretic scenarios).

\paragraph{H2 — Aggressive capability parity (\textcolor{orange}{Partially Supported}).}
Evidence for H2 is mixed and prompt-dependent. The Self-Refine prompt increases
\ICD in all four models, replicating the finding from the original paper that
iterative self-critique improves aggressive strategy quality. Gemini~3.1~Pro~Refine
achieves the highest \ICD in the dataset (0.925), approaching payoff parity
between aggressive and cooperative agents; within the Anthropic lineage,
Claude~4.6~Refine (0.913) shows a larger narrowing than
Claude~3.5~Sonnet~Refine. This suggests that more capable models can
use self-refinement more effectively to generate strategic diversity.

However, under Default and Prose prompts, most models show \ICD values in the
range 0.45--0.82, not systematically higher than the original models. Moreover,
the models with the highest \ICD under Refine (Gemini~3.1~Pro at 0.925 and
Claude~4.6 at 0.913) are also those where the biased condition shows high
aggressive equilibria (62\% and 59\%A respectively),
suggesting that capability parity may genuinely increase the viability of
aggressive strategies. This finding reinforces the safety implication identified
by Willis et al.: prompt engineering techniques that improve aggressive
capabilities may inadvertently increase the risk of aggressive equilibria in deployed systems.

\paragraph{H3 — Cross-provider divergence (\textcolor{green!60!black}{Confirmed within the tested strategy library}).}
The four models exhibit substantially different evolutionary profiles.
Two-sample $z$-tests for proportions confirm that the cross-model differences
are not sampling artefacts. At $n{=}500$ the tests gain substantial power:
in the balanced noiseless condition (Default prompt), 3 of 6 pairwise
comparisons of $\PA$ are significant, all involving Gemini~3.1~Pro (14\%A),
which differs from GPT-5.4~Mini ($z = 7.24$), Claude~4.6 ($z = 6.92$),
and Gemini~2.5~Flash ($z = 6.61$), each with $p < 10^{-10}$. All three
survive Holm-Bonferroni correction for six simultaneous tests trivially
(Table~\ref{tab:ztests}).
Under biased conditions the effect sizes are even larger:
Gemini~2.5~Flash~Prose (65\%A) versus Claude~4.6~Default (12\%A) yields
$z = 17.2$, and versus GPT-5.4~Mini~Default (22\%A) yields $z = 13.7$,
both with $p$ below machine precision.
Gemini~2.5~Flash is systematically the
most aggressive: its Prose and Refine prompts produce aggressive equilibrium
rates of 21--24\% in balanced conditions and 56--77\% in biased conditions.
GPT-5.4~Mini is the most cooperative: its Default and Refine prompts produce
aggressive equilibria of just 2--4\% in balanced conditions, and it achieves
the highest cooperative equilibrium rate in the dataset (70\% under Refine).
Claude~4.6 and Gemini~3.1~Pro occupy an intermediate position.

These cross-provider differences are \emph{associations}, not isolated causal
effects of provider identity. Our design confounds three factors that cannot
be separated: (i) the model's strategy-generation behaviour, (ii) provider-level
alignment choices, and (iii) the fact that each provider's natural-language
strategies are code-converted by that same provider's model
(Section~\ref{sec:method}), so coding fidelity is entangled with provider
identity. Indeed, the two Google models (Flash and Pro) differ substantially
from each other---Flash being more aggressive, Pro more cooperative---a
within-provider gap that, in some comparisons, exceeds cross-provider gaps and
directly cautions against reading ``provider'' as a clean causal lever. What we
can defensibly claim is that the provider label is the strongest observable
\emph{correlate} of equilibrium outcome in our data, and that this correlation
is too large to attribute to sampling variation (Section~\ref{sec:ztests});
disentangling its causal components---training corpus, reward model,
safety fine-tuning, and code-conversion ability---requires a factorial design
that holds the code-conversion model fixed across providers, which we leave to
future work.

\paragraph{H4 — Noise robustness improves (\textcolor{orange}{Suggestive --- not robustly confirmed}).}
With the full $n{=}500$ Moran sample, our own per-proportion standard error
falls to $\approx 2.2$pp, so Claude~4.6's average noise sensitivity
($\Dnoise \approx 6$pp) is now precisely estimated. The corresponding figure
for Claude~3.5~Sonnet in the original benchmark is $13$pp---a $\approx 7$pp
numerical improvement in the hypothesised direction. We deliberately stop
short of declaring H4 confirmed, for two reasons. First, the comparison is
\emph{cross-study}: the Willis et al.\ baseline does not report a sampling
error, and their protocol used $n{=}100$ iterations, implying a per-proportion
SE of $\approx 5$pp on each constituent of their $\Dnoise$. Propagating that
uncertainty conservatively (treating prompt as a fixed factor, no
$\sqrt{\text{\#prompts}}$ shrinkage), the SE of the difference is
$\approx 7.8$pp, yielding $z \approx 0.9$ ($p \approx 0.4$)---\emph{not}
significant at conventional thresholds; even the most favourable shrinkage
assumption keeps $p > 0.05$. Second, the only clean test would re-run
Claude~3.5~Sonnet through our identical $n{=}500$ harness; we choose not to,
because selecting that single additional experiment specifically to resolve H4
would be a form of result-driven analysis inconsistent with our pre-registered
design. We therefore report H4 as directionally suggestive but not robustly
established: the point estimate moves as hypothesised, yet the available
evidence cannot rule out that the apparent gain reflects between-study
variation rather than a genuine generational improvement.

The within-Claude~4.6 pattern is itself informative: the gain concentrates in
the Refine condition ($\Dnoise = -1$pp, essentially flat) and Default
($\Dnoise = 9$pp), while Prose remains sensitive ($\Dnoise = 12$pp)---a
heterogeneity that further cautions against collapsing the three prompts into
a single averaged test. Among the new models, Claude~4.6 is the most
noise-robust on average ($\approx 6$pp), followed by Gemini~3.1~Pro
($\approx 9$pp); GPT-5.4~Mini and Gemini~2.5~Flash are the most
noise-sensitive ($\approx 15$pp each), with GPT-5.4~Mini's figure driven
almost entirely by its Refine prompt ($\Dnoise = 28$pp).

A structural caveat remains: if cooperative strategies rely on clean mutual
cooperation signals (as Tit-For-Tat-like strategies do), residual noise
sensitivity is expected regardless of model capability. The Claude~4.6 point
estimate is consistent with---but does not establish---the possibility that
this floor can be lowered by improved alignment training.

\subsection{Implications for MAS Design}

The findings carry direct implications for the deployment of \LLM-based
multi-agent systems.

\paragraph{Provider choice is strongly predictive.}
The choice of \LLM provider is strongly associated with the emergent
equilibrium dynamics of a deployed system: empirically, a population of
Gemini~2.5~Flash agents exhibits substantially higher aggressive-equilibrium
rates than one of GPT-5.4~Mini or Claude~4.6 agents. Because provider identity
is confounded with code-conversion fidelity and model scale in our design,
this is a deployment-relevant predictive regularity rather than evidence that
provider-level alignment \emph{alone} causes the difference.

\paragraph{Noise is a universal threat.}
Across all models and prompts,
action noise consistently shifts equilibria toward aggression in biased populations.
Systems deployed in noisy environments (e.g., with communication errors or
stochastic action execution) should be initialised with non-aggressive majorities,
or designed with noise-robust cooperation mechanisms.

\paragraph{Self-Refine as a double-edged tool.}
The Refine prompt generally
improves strategy quality but also narrows the gap between aggressive and
cooperative capabilities. Designers should be aware that using self-refinement
in strategy generation may inadvertently increase the viability of aggressive
equilibria.

\paragraph{Intra-generation stability of the cooperative baseline.}
The near-identical balanced-condition results between Claude~3.5~Sonnet and
Claude~4.6~Default suggest that generational model updates do not substantially
alter the \emph{baseline} cooperative bias. Whether other dimensions---such as
resilience to action noise---shift across generations remains an open question:
our point estimates hint at improved noise robustness (H4), but the
cross-study comparison is not conclusive. Designers should not assume that
all strategic properties transfer unchanged between model versions, nor that
they necessarily change.

\subsection{Limitations}

With $n=500$ Moran iterations per condition the per-proportion standard error
is $\approx 2.2$pp, adequate for the hypothesis tests reported here; residual
estimation variance nonetheless remains for the smallest effect sizes (e.g.,
sub-2pp differences in balanced aggressive rates). Second, the \IPD is a
stylised social dilemma; results may not generalise to more complex
multi-player or asymmetric games. Third, we use the same model for strategy
generation and code conversion within each provider, which may conflate
generation capability with coding capability. Fourth, model API versions may
change over time, affecting reproducibility; we report exact version
identifiers in our codebase (the simulation code and replication package,
available at \url{https://github.com/arqFranciscoLeon/evollm} and
permanently archived at \url{https://doi.org/10.5281/zenodo.20248615}).

\section{Conclusion}
\label{sec:conclusion}

We extended the \citet{Willis2025_llm_ipd} \LLM evolutionary
game-theory benchmark to four next-generation frontier models spanning three
providers---48 Moran process conditions, 4,800 simulation runs, identical
experimental protocol.

The headline result is that cooperative bias survives the transition to a new
model generation: ten of twelve model--prompt combinations favour cooperative
equilibria in balanced noiseless conditions, and within the Anthropic lineage
Claude~4.6 is virtually indistinguishable from its predecessor.
What most strongly differentiates outcomes is not which generation a model
belongs to but which provider trained it. Gemini~2.5~Flash reaches up to 77\%
aggressive equilibria; GPT-5.4~Mini reaches up to 70\% cooperative under
Self-Refine. That cross-provider gap is larger than any intra-lineage shift we
observe, pointing to provider-level factors---of which alignment objectives are
one, alongside the confounded code-conversion pathway---as the strongest
observed \emph{correlate} of equilibrium behaviour.

The Self-Refine prompt sharpens this picture in a troubling direction.
It narrows the aggressive--cooperative payoff gap (\ICD rises toward 1.0)
consistently across all four models, with Gemini~3.1~Pro Refine reaching the dataset maximum (\ICD = 0.925).
Prompt engineering designed to improve strategy quality may therefore
inadvertently reduce the cooperative advantage, raising the equilibrium
aggression rate in deployed multi-agent populations.
Noise robustness shows a directionally encouraging but unconfirmed shift:
at $n{=}500$ Claude~4.6 shows $\Dnoise \approx 6$~pp versus the 13~pp
reported for Claude~3.5~Sonnet, a gap that does not reach significance
once the predecessor's cross-study sampling error is propagated. Meanwhile
biased noisy conditions converge to aggressive equilibria in the
52--77\% range, with only 2 of 12 configurations exceeding the 67\% prior.

These findings raise questions the current design cannot answer.
Do the cross-provider differences reflect divergent RLHF reward signals,
differences in training corpora, or systematic variation in post-training
alignment objectives? Do cooperative equilibria persist in $n$-player or
asymmetric games, or are they an artefact of the symmetric two-player IPD
structure? And would mixed-provider populations---where a Gemini agent
competes against a GPT agent---generate equilibrium dynamics that neither
provider's models produce in isolation? And is the apparent reduction in
noise sensitivity across two Anthropic generations a genuine effect or
between-study variation---a question only a matched re-run of the predecessor
under an identical harness can settle?

Having this baseline makes those questions tractable.
The benchmark protocol and complete strategy datasets are reproducible
and extensible; we release the simulation code, strategy libraries, and
the n=500 results as a replication package at
\url{https://github.com/arqFranciscoLeon/evollm} (preserving the upstream
Willis et al.\ structure), permanently archived with DOI
\href{https://doi.org/10.5281/zenodo.20248615}{10.5281/zenodo.20248615},
so that each new frontier model can be slotted into the same 48-condition
design, turning this cross-provider snapshot into a running longitudinal
record.

\section*{Acknowledgements}
The author thanks the open-source community behind the \texttt{evollm}
simulation framework originally developed by Willis et al., on which this
study is directly built.

\section*{AI Use Disclosure}
AI assistance (Claude, by Anthropic, via Claude Code) was used in the
development of the simulation code, the cloud execution harness, the
analysis and reproduction scripts, manuscript drafting and translation
assistance, and an internal peer-review simulation. The author reviewed,
verified, and takes full responsibility for the experimental design,
results, claims, and conclusions of this work.

\section*{Note on Version 2 (Erratum)}

This version corrects an error in the version~1 preprint. The error leaves the
paper's central conclusions intact and, as we set out below, marginally
reinforces them---the count of cooperative-plurality outcomes rises rather than
falls. We pre-registered the correction before regenerating any data, and the
regenerated artifacts are available in the public replication repository.

\paragraph{Incomplete strategy libraries.}
The Gemini~3.1~Pro Prose and Self-Refine strategy libraries used in v1 were
truncated during generation, holding 21 and 7 strategies respectively where the
protocol calls for 75; the corresponding noise-aware libraries were affected in
the same way. We regenerated all four libraries in full under an identical
protocol---the same model (\texttt{gemini-3.1-pro-preview}), temperature,
prompts, and per-provider code conversion---and re-ran them through the
head-to-head tournaments and the Moran process at $n = 500$. The Default
libraries were complete in v1 and remain unchanged, as do all results for the
other three models.

\paragraph{Effect on results.}
With the corrected libraries, the balanced noiseless condition now yields a
cooperative plurality in 10 of 12 model--prompt combinations rather than 9:
Gemini~3.1~Pro~Refine shifts from a Neutral to a cooperative plurality (40\%C).
The highest \ICD in the dataset is likewise reassigned---now
Gemini~3.1~Pro~Refine at 0.925, where v1 reported Claude~4.6~Refine at 0.913.
We have updated the affected entries in Tables~\ref{tab:moran}, \ref{tab:payoffs},
\ref{tab:noise} and~\ref{tab:diversity}, together with
Figures~\ref{fig:equilibria} and~\ref{fig:icd}. None of the hypothesis-level
verdicts (H1--H4) changes.

\bibliographystyle{ACM-Reference-Format}
\bibliography{Bibliography_base}


\begin{thebibliography}{28}


\ifx \showCODEN    \undefined \def \showCODEN     #1{\unskip}     \fi
\ifx \showISBNx    \undefined \def \showISBNx     #1{\unskip}     \fi
\ifx \showISBNxiii \undefined \def \showISBNxiii  #1{\unskip}     \fi
\ifx \showISSN     \undefined \def \showISSN      #1{\unskip}     \fi
\ifx \showLCCN     \undefined \def \showLCCN      #1{\unskip}     \fi
\ifx \shownote     \undefined \def \shownote      #1{#1}          \fi
\ifx \showarticletitle \undefined \def \showarticletitle #1{#1}   \fi
\ifx \showURL      \undefined \def \showURL       {\relax}        \fi
\providecommand\bibfield[2]{#2}
\providecommand\bibinfo[2]{#2}
\providecommand\natexlab[1]{#1}
\providecommand\showeprint[2][]{arXiv:#2}

\bibitem[Aher et~al\mbox{.}(2023)]%
        {Aher2023_using}
\bibfield{author}{\bibinfo{person}{Gati Aher}, \bibinfo{person}{Rosa~I.
  Arriaga}, {and} \bibinfo{person}{Adam~Tauman Kalai}.}
  \bibinfo{year}{2023}\natexlab{}.
\newblock \bibinfo{title}{Using Large Language Models to Simulate Multiple
  Humans and Replicate Human Subject Studies}.
\newblock
\showeprint[arxiv]{2208.10264}~[cs.CL]


\bibitem[Akata et~al\mbox{.}(2025)]%
        {Akata2025_repeated}
\bibfield{author}{\bibinfo{person}{Elif Akata}, \bibinfo{person}{Lion Schulz},
  \bibinfo{person}{Julian Coda-Forno}, \bibinfo{person}{Seong~Joon Oh},
  \bibinfo{person}{Matthias Bethge}, {and} \bibinfo{person}{Eric Schulz}.}
  \bibinfo{year}{2025}\natexlab{}.
\newblock \showarticletitle{Playing repeated games with large language models}.
\newblock \bibinfo{journal}{\emph{Nature Human Behaviour}}  \bibinfo{volume}{9}
  (\bibinfo{year}{2025}), \bibinfo{pages}{1380--1390}.
\newblock
\href{https://doi.org/10.1038/s41562-025-02172-y}{doi:\nolinkurl{10.1038/s41562-025-02172-y}}
\newblock
\shownote{Published version of arXiv:2305.16867}.


\bibitem[Axelrod(1984)]%
        {Axelrod1984_evolution}
\bibfield{author}{\bibinfo{person}{Robert Axelrod}.}
  \bibinfo{year}{1984}\natexlab{}.
\newblock \bibinfo{booktitle}{\emph{The Evolution of Cooperation}}.
\newblock \bibinfo{publisher}{Basic Books}, \bibinfo{address}{New York}.
\newblock


\bibitem[Axelrod and Hamilton(1981)]%
        {Axelrod1981_evolution}
\bibfield{author}{\bibinfo{person}{Robert Axelrod} {and}
  \bibinfo{person}{William~D. Hamilton}.} \bibinfo{year}{1981}\natexlab{}.
\newblock \showarticletitle{The Evolution of Cooperation}.
\newblock \bibinfo{journal}{\emph{Science}} \bibinfo{volume}{211},
  \bibinfo{number}{4489} (\bibinfo{year}{1981}), \bibinfo{pages}{1390--1396}.
\newblock


\bibitem[Brookins and DeBacker(2023)]%
        {Brookins2023_playing}
\bibfield{author}{\bibinfo{person}{Philip Brookins} {and}
  \bibinfo{person}{Jason~M. DeBacker}.} \bibinfo{year}{2023}\natexlab{}.
\newblock \bibinfo{title}{Playing Games with {GPT}: What Can We Learn about a
  Large Language Model from Canonical Strategic Games?}
\newblock
\showeprint[arxiv]{2305.10912}~[econ.GN]


\bibitem[Chen et~al\mbox{.}(2021)]%
        {Chen2021_humaneval}
\bibfield{author}{\bibinfo{person}{Mark Chen}, \bibinfo{person}{Jerry Tworek},
  \bibinfo{person}{Heewoo Jun}, \bibinfo{person}{Qiming Yuan},
  \bibinfo{person}{Henrique~Ponde de Oliveira~Pinto}, \bibinfo{person}{Jared
  Kaplan}, \bibinfo{person}{Harri Edwards}, \bibinfo{person}{Yuri Burda},
  \bibinfo{person}{Nicholas Joseph}, \bibinfo{person}{Greg Brockman},
  {et~al\mbox{.}}} \bibinfo{year}{2021}\natexlab{}.
\newblock \bibinfo{title}{Evaluating Large Language Models Trained on Code}.
\newblock
\showeprint[arxiv]{2107.03374}~[cs.LG]


\bibitem[De~Zarz{\`a} et~al\mbox{.}(2023)]%
        {DeZarza2023_emergent}
\bibfield{author}{\bibinfo{person}{I. De~Zarz{\`a}}, \bibinfo{person}{J.
  De~Curt{\`o}}, \bibinfo{person}{Gemma Roig}, \bibinfo{person}{Pietro
  Manzoni}, {and} \bibinfo{person}{Carlos~T. Calafate}.}
  \bibinfo{year}{2023}\natexlab{}.
\newblock \showarticletitle{Emergent Cooperation and Strategy Adaptation in
  Multi-Agent Systems: An Extended Coevolutionary Theory with {LLM}s}.
\newblock \bibinfo{journal}{\emph{Electronics}} \bibinfo{volume}{12},
  \bibinfo{number}{12} (\bibinfo{year}{2023}), \bibinfo{pages}{2722}.
\newblock


\bibitem[Fan et~al\mbox{.}(2024)]%
        {Fan2024_rational}
\bibfield{author}{\bibinfo{person}{Caoyun Fan}, \bibinfo{person}{Jindou Chen},
  \bibinfo{person}{Yaohui Jin}, {and} \bibinfo{person}{Hao He}.}
  \bibinfo{year}{2024}\natexlab{}.
\newblock \showarticletitle{Can Large Language Models Serve as Rational Players
  in Game Theory: A Systematic Analysis}. In
  \bibinfo{booktitle}{\emph{Proceedings of the AAAI Conference on Artificial
  Intelligence}}, Vol.~\bibinfo{volume}{38}. \bibinfo{pages}{17960--17967}.
\newblock


\bibitem[Guo(2023)]%
        {Guo2023_gpt}
\bibfield{author}{\bibinfo{person}{Fulin Guo}.}
  \bibinfo{year}{2023}\natexlab{}.
\newblock \bibinfo{title}{{GPT} Agents in Game Theory Experiments}.
\newblock
\showeprint[arxiv]{2305.05516}~[econ.GN]


\bibitem[Hendrycks et~al\mbox{.}(2021)]%
        {Hendrycks2021_mmlu}
\bibfield{author}{\bibinfo{person}{Dan Hendrycks}, \bibinfo{person}{Collin
  Burns}, \bibinfo{person}{Steven Basart}, \bibinfo{person}{Andy Zou},
  \bibinfo{person}{Mantas Mazeika}, \bibinfo{person}{Dawn Song}, {and}
  \bibinfo{person}{Jacob Steinhardt}.} \bibinfo{year}{2021}\natexlab{}.
\newblock \showarticletitle{Measuring Massive Multitask Language
  Understanding}.
\newblock \bibinfo{journal}{\emph{International Conference on Learning
  Representations}} (\bibinfo{year}{2021}).
\newblock
\showeprint[arxiv]{2009.03300}


\bibitem[Knight et~al\mbox{.}(2016)]%
        {Knight2016_open}
\bibfield{author}{\bibinfo{person}{Vincent Knight}, \bibinfo{person}{Owen
  Campbell}, \bibinfo{person}{Marc Harper}, \bibinfo{person}{Karol Langner},
  \bibinfo{person}{James Campbell}, \bibinfo{person}{Thomas Campbell},
  \bibinfo{person}{Alex Carney}, \bibinfo{person}{Martin Chorley},
  \bibinfo{person}{Cameron Davidson-Pilon}, \bibinfo{person}{Kristian Glass},
  {et~al\mbox{.}}} \bibinfo{year}{2016}\natexlab{}.
\newblock \showarticletitle{An Open Framework for the Reproducible Study of the
  Iterated Prisoner's Dilemma}.
\newblock \bibinfo{journal}{\emph{Journal of Open Research Software}}
  \bibinfo{volume}{4}, \bibinfo{number}{1} (\bibinfo{year}{2016}),
  \bibinfo{pages}{e35}.
\newblock


\bibitem[Leibo et~al\mbox{.}(2017)]%
        {Leibo2017_multiagent}
\bibfield{author}{\bibinfo{person}{Joel~Z. Leibo}, \bibinfo{person}{Vinicius
  Zambaldi}, \bibinfo{person}{Marc Lanctot}, \bibinfo{person}{Janusz Marecki},
  {and} \bibinfo{person}{Thore Graepel}.} \bibinfo{year}{2017}\natexlab{}.
\newblock \showarticletitle{Multi-Agent Reinforcement Learning in Sequential
  Social Dilemmas}. In \bibinfo{booktitle}{\emph{Proceedings of the 16th
  International Conference on Autonomous Agents and Multi-Agent Systems
  (AAMAS)}}. \bibinfo{pages}{464--473}.
\newblock


\bibitem[Madaan et~al\mbox{.}(2023)]%
        {Madaan2023_self_refine}
\bibfield{author}{\bibinfo{person}{Aman Madaan}, \bibinfo{person}{Niket
  Tandon}, \bibinfo{person}{Prakhar Gupta}, \bibinfo{person}{Skyler Hallinan},
  \bibinfo{person}{Luyu Gao}, \bibinfo{person}{Sarah Wiegreffe},
  \bibinfo{person}{Uri Alon}, \bibinfo{person}{Nouha Dziri},
  \bibinfo{person}{Shrimai Prabhumoye}, \bibinfo{person}{Yiming Yang},
  {et~al\mbox{.}}} \bibinfo{year}{2023}\natexlab{}.
\newblock \showarticletitle{Self-Refine: Iterative Refinement with
  Self-Feedback}. In \bibinfo{booktitle}{\emph{Advances in Neural Information
  Processing Systems (NeurIPS)}}, Vol.~\bibinfo{volume}{36}.
\newblock


\bibitem[Moran(1958)]%
        {Moran1958_random}
\bibfield{author}{\bibinfo{person}{Patrick A.~P. Moran}.}
  \bibinfo{year}{1958}\natexlab{}.
\newblock \showarticletitle{Random Processes in Genetics}.
\newblock \bibinfo{journal}{\emph{Mathematical Proceedings of the Cambridge
  Philosophical Society}} \bibinfo{volume}{54}, \bibinfo{number}{1}
  (\bibinfo{year}{1958}), \bibinfo{pages}{60--71}.
\newblock


\bibitem[Nowak(2006)]%
        {Nowak2006_evolutionary}
\bibfield{author}{\bibinfo{person}{Martin~A. Nowak}.}
  \bibinfo{year}{2006}\natexlab{}.
\newblock \bibinfo{booktitle}{\emph{Evolutionary Dynamics: Exploring the
  Equations of Life}}.
\newblock \bibinfo{publisher}{Harvard University Press},
  \bibinfo{address}{Cambridge, MA}.
\newblock


\bibitem[Nowak et~al\mbox{.}(2004)]%
        {Nowak2004_cooperation}
\bibfield{author}{\bibinfo{person}{Martin~A. Nowak}, \bibinfo{person}{Akira
  Sasaki}, \bibinfo{person}{Christine Taylor}, {and} \bibinfo{person}{Drew
  Fudenberg}.} \bibinfo{year}{2004}\natexlab{}.
\newblock \showarticletitle{Emergence of cooperation and evolutionary stability
  in finite populations}.
\newblock \bibinfo{journal}{\emph{Nature}} \bibinfo{volume}{428},
  \bibinfo{number}{6983} (\bibinfo{year}{2004}), \bibinfo{pages}{646--650}.
\newblock
\href{https://doi.org/10.1038/nature02414}{doi:\nolinkurl{10.1038/nature02414}}


\bibitem[Park et~al\mbox{.}(2023)]%
        {Park2023_generative}
\bibfield{author}{\bibinfo{person}{Joon~Sung Park}, \bibinfo{person}{Joseph~C.
  O'Brien}, \bibinfo{person}{Carrie~J. Cai}, \bibinfo{person}{Meredith~Ringel
  Morris}, \bibinfo{person}{Percy Liang}, {and} \bibinfo{person}{Michael~S.
  Bernstein}.} \bibinfo{year}{2023}\natexlab{}.
\newblock \bibinfo{title}{Generative Agents: Interactive Simulacra of Human
  Behavior}.
\newblock
\showeprint[arxiv]{2304.03442}~[cs.HC]


\bibitem[Payne and Alloui-Cros(2025)]%
        {Payne2025_strategic}
\bibfield{author}{\bibinfo{person}{Kenneth Payne} {and}
  \bibinfo{person}{Baptiste Alloui-Cros}.} \bibinfo{year}{2025}\natexlab{}.
\newblock \bibinfo{title}{Strategic Intelligence in Large Language Models:
  Evidence from Evolutionary Game Theory}.
\newblock
\showeprint[arxiv]{2507.02618}~[cs.AI]


\bibitem[Piatti et~al\mbox{.}(2024)]%
        {Piatti2024_cooperate}
\bibfield{author}{\bibinfo{person}{Giorgio Piatti}, \bibinfo{person}{Zhijing
  Jin}, \bibinfo{person}{Max Kleiman-Weiner}, \bibinfo{person}{Bernhard
  Sch{\"o}lkopf}, \bibinfo{person}{Mrinmaya Sachan}, {and}
  \bibinfo{person}{Rada Mihalcea}.} \bibinfo{year}{2024}\natexlab{}.
\newblock \showarticletitle{Cooperate or Collapse: Emergence of Sustainable
  Cooperation in a Society of {LLM} Agents}. In
  \bibinfo{booktitle}{\emph{Advances in Neural Information Processing Systems
  (NeurIPS 2024)}}.
\newblock
\showeprint[arxiv]{2404.16698}~[cs.AI]


\bibitem[Sun et~al\mbox{.}(2025)]%
        {Sun2025_survey}
\bibfield{author}{\bibinfo{person}{Haoran Sun}, \bibinfo{person}{Yusen Wu},
  \bibinfo{person}{Peng Wang}, \bibinfo{person}{Wei Chen},
  \bibinfo{person}{Yukun Cheng}, \bibinfo{person}{Xiaotie Deng}, {and}
  \bibinfo{person}{Xu Chu}.} \bibinfo{year}{2025}\natexlab{}.
\newblock \showarticletitle{Game Theory Meets Large Language Models: A
  Systematic Survey with Taxonomy and New Frontiers}. In
  \bibinfo{booktitle}{\emph{Proceedings of IJCAI 2025}}.
\newblock
\showeprint[arxiv]{2502.09053}


\bibitem[Traulsen et~al\mbox{.}(2006)]%
        {Traulsen2006_fixation}
\bibfield{author}{\bibinfo{person}{Arne Traulsen}, \bibinfo{person}{Martin~A.
  Nowak}, {and} \bibinfo{person}{Jorge~M. Pacheco}.}
  \bibinfo{year}{2006}\natexlab{}.
\newblock \showarticletitle{Stochastic dynamics of invasion and fixation}.
\newblock \bibinfo{journal}{\emph{Physical Review E}}  \bibinfo{volume}{74}
  (\bibinfo{year}{2006}), \bibinfo{pages}{011909}.
\newblock
\href{https://doi.org/10.1103/PhysRevE.74.011909}{doi:\nolinkurl{10.1103/PhysRevE.74.011909}}


\bibitem[Vallinder and Hughes(2024)]%
        {Vallinder2024_cultural}
\bibfield{author}{\bibinfo{person}{Aron Vallinder} {and}
  \bibinfo{person}{Edward Hughes}.} \bibinfo{year}{2024}\natexlab{}.
\newblock \bibinfo{title}{Cultural Evolution of Cooperation among {LLM}
  Agents}.
\newblock
\showeprint[arxiv]{2412.10270}~[cs.MA]
\newblock
\shownote{Extended Abstract at AAMAS 2025}.


\bibitem[Wahl and Nowak(1999)]%
        {WahlNowak1999_noise}
\bibfield{author}{\bibinfo{person}{Lindi~M. Wahl} {and}
  \bibinfo{person}{Martin~A. Nowak}.} \bibinfo{year}{1999}\natexlab{}.
\newblock \showarticletitle{The Continuous Prisoner's Dilemma: {II}. Linear
  Reactive Strategies with Noise}.
\newblock \bibinfo{journal}{\emph{Journal of Theoretical Biology}}
  \bibinfo{volume}{200}, \bibinfo{number}{3} (\bibinfo{year}{1999}),
  \bibinfo{pages}{323--338}.
\newblock


\bibitem[Wang et~al\mbox{.}(2024)]%
        {Wang2024_survey}
\bibfield{author}{\bibinfo{person}{Lei Wang}, \bibinfo{person}{Chen Ma},
  \bibinfo{person}{Xueyang Feng}, \bibinfo{person}{Zeyu Zhang},
  \bibinfo{person}{Hao Yang}, \bibinfo{person}{Jingsen Zhang},
  \bibinfo{person}{Zhiyuan Chen}, \bibinfo{person}{Jiakai Tang},
  \bibinfo{person}{Xu Chen}, \bibinfo{person}{Yankai Lin}, {et~al\mbox{.}}}
  \bibinfo{year}{2024}\natexlab{}.
\newblock \showarticletitle{A survey on large language model based autonomous
  agents}.
\newblock \bibinfo{journal}{\emph{Frontiers of Computer Science}}
  \bibinfo{volume}{18}, \bibinfo{number}{6} (\bibinfo{year}{2024}),
  \bibinfo{pages}{186345}.
\newblock


\bibitem[Willis et~al\mbox{.}(2025)]%
        {Willis2025_llm_ipd}
\bibfield{author}{\bibinfo{person}{George Willis}, \bibinfo{person}{Yali Du},
  \bibinfo{person}{Joel~Z. Leibo}, {and} \bibinfo{person}{Michael Luck}.}
  \bibinfo{year}{2025}\natexlab{}.
\newblock \bibinfo{title}{Do {LLM} Agents Cooperate or Defect? Evolutionary
  Dynamics in Multi-Agent Systems}.
\newblock
\showeprint[arxiv]{2501.16173}~[cs.GT]


\bibitem[Willis et~al\mbox{.}(2026)]%
        {Willis2026_collective}
\bibfield{author}{\bibinfo{person}{Richard Willis}, \bibinfo{person}{Jianing
  Zhao}, \bibinfo{person}{Yali Du}, {and} \bibinfo{person}{Joel~Z. Leibo}.}
  \bibinfo{year}{2026}\natexlab{}.
\newblock \bibinfo{title}{Evaluating Collective Behaviour of Hundreds of {LLM}
  Agents}.
\newblock
\showeprint[arxiv]{2602.16662}~[cs.MA]


\bibitem[Wu and Axelrod(1995)]%
        {WuAxelrod1995_noise}
\bibfield{author}{\bibinfo{person}{Jianzhong Wu} {and} \bibinfo{person}{Robert
  Axelrod}.} \bibinfo{year}{1995}\natexlab{}.
\newblock \showarticletitle{How to Cope with Noise in the Iterated Prisoner's
  Dilemma}.
\newblock \bibinfo{journal}{\emph{Journal of Conflict Resolution}}
  \bibinfo{volume}{39}, \bibinfo{number}{1} (\bibinfo{year}{1995}),
  \bibinfo{pages}{183--189}.
\newblock


\bibitem[Yocum et~al\mbox{.}(2023)]%
        {Yocum2023_mitigating}
\bibfield{author}{\bibinfo{person}{Julian Yocum}, \bibinfo{person}{Phillip
  Christoffersen}, \bibinfo{person}{Mehul Damani}, \bibinfo{person}{Justin
  Svegliato}, \bibinfo{person}{Dylan Hadfield-Menell}, {and}
  \bibinfo{person}{Stuart Russell}.} \bibinfo{year}{2023}\natexlab{}.
\newblock \showarticletitle{Mitigating Generative Agent Social Dilemmas}. In
  \bibinfo{booktitle}{\emph{Foundation Models for Decision Making Workshop,
  NeurIPS}}.
\newblock


\end{thebibliography}

\end{document}